\documentclass[8.5pt,twoside,twocolumn]{article}
\usepackage[super,sort&compress,comma]{natbib}

\usepackage{times} %,mathptmx
\usepackage{sectsty}
\usepackage{balance}

\usepackage{graphicx} %eps figures can be used instead
\usepackage{lastpage}
\usepackage[format=plain,singlelinecheck=false,font=small,labelfont=bf,labelsep=space]{caption} %justification=raggedright,
\usepackage{fancyhdr}
\usepackage{color}
\usepackage{amsmath}

\begin{document}

\thispagestyle{plain}
\fancypagestyle{plain}{
\fancyhead[L]{\includegraphics[height=8pt]{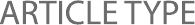}}
\fancyhead[C]{\hspace{-1cm}\includegraphics[height=20pt]{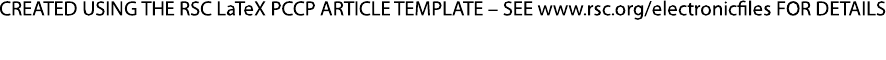}}
\fancyhead[R]{\includegraphics[height=10pt]{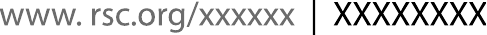}\vspace{-0.2cm}}
\renewcommand{\headrulewidth}{1pt}}
\renewcommand{\thefootnote}{\fnsymbol{footnote}}
\renewcommand\footnoterule{\vspace*{1pt}%
\hrule width 3.4in height 0.4pt \vspace*{5pt}}
\setcounter{secnumdepth}{5}

\makeatletter
\def\subsubsection{\@startsection{subsubsection}{3}{10pt}{-1.25ex plus -1ex minus -.1ex}{0ex plus 0ex}{\normalsize\bf}}
\def\paragraph{\@startsection{paragraph}{4}{10pt}{-1.25ex plus -1ex minus -.1ex}{0ex plus 0ex}{\normalsize\textit}}
\renewcommand\@biblabel[1]{#1}
\renewcommand\@makefntext[1]%
{\noindent\makebox[0pt][r]{\@thefnmark\,}#1}
\makeatother
\renewcommand{\figurename}{\small{Fig.}~}
\sectionfont{\large}
\subsectionfont{\normalsize}

\fancyfoot{}
\fancyfoot[LO,RE]{\vspace{-7pt}\includegraphics[height=9pt]{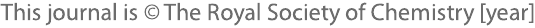}}
\fancyfoot[CO]{\vspace{-7.2pt}\hspace{12.2cm}\includegraphics{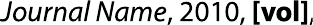}}
\fancyfoot[CE]{\vspace{-7.5pt}\hspace{-13.5cm}\includegraphics{RF}}
\fancyfoot[RO]{\footnotesize{\sffamily{1--\pageref{LastPage} ~\textbar  \hspace{2pt}\thepage}}}
\fancyfoot[LE]{\footnotesize{\sffamily{\thepage~\textbar\hspace{3.45cm} 1--\pageref{LastPage}}}}
\fancyhead{}
\renewcommand{\headrulewidth}{1pt}
\renewcommand{\footrulewidth}{1pt}
\setlength{\arrayrulewidth}{1pt}
\setlength{\columnsep}{6.5mm}
\setlength\bibsep{1pt}

%%%%%%%%%%%%%%%%%%%%%%%%%%%%%%%%%%%%%%%%%%%%%%%%%%%%%%%%% ourdef
\def\mycirc{{\large $\circ$}}
\definecolor{darkgreen}{RGB}{0,139,0}
\definecolor{turqoise}{RGB}{64,224,208}
\definecolor{brown}{RGB}{210,105,30}
\def\myeightstar{\protect\makebox[0pt][l]{+}$\times$}
\definecolor{b}{rgb}{0,0,1.0}
\definecolor{r}{rgb}{1,0,0}
\definecolor{g}{rgb}{0,1,0}
\newcommand{\BT}[1]{{\color{b}{#1}}}
\newcommand{\OK}[1]{{\color{darkgreen}{#1}}}
\newcommand{\Quest}[1]{{\color{r}{#1}}}
\newcommand{\MOD}[1]{{\color{magenta}{#1}}}
\newcommand{\MM}[1]{{\color{brown}{#1}}}
\def\T{\textit{\textbf{T}}}
%%%%%%%%%%%%%%%%%%%%%%%%%%%%%%%%%%%%%%%%%%%%%%%%%%% end ourdef

\twocolumn[
  \begin{@twocolumnfalse}
\noindent\LARGE{\textbf{Effects of grain shape on packing and dilatancy of sheared granular materials}}
\vspace{0.6cm}

\noindent\large{\textbf{Sandra Wegner,$^{\ast}$\textit{$^{a}$} Ralf Stannarius,\textit{$^{a}$} Axel Boese,\textit{$^{b}$}
Georg Rose,\textit{$^{b}$} Bal\'azs Szab\'o,\textit{$^{c}$} Ell\'ak Somfai,\textit{$^{c}$}  and 
Tam\'as B\"orzs\"onyi,\textit{$^{c}$}}}\vspace{0.5cm}

\noindent\textit{\small{\textbf{Received Xth XXXXXXXXXX 20XX, Accepted Xth XXXXXXXXX 20XX\newline
First published on the web Xth XXXXXXXXXX 200X}}}

\noindent \textbf{\small{DOI: 10.1039/b000000x}}
\vspace{0.6cm}
%Please do not change this text.

\noindent \normalsize{
Granular material exposed to shear shows a variety of unique phenomena: Reynolds dilatancy, positional order and orientational order effects may compete in the shear zone. We study granular packings consisting of macroscopic prolate, oblate and spherical grains and compare their behaviour. X-ray tomography is used to determine the particle positions and orientations in a cylindrical split bottom shear cell. Packing densities and the arrangements of individual particles in the shear zone are evaluated. For anisometric particles, we observe the competition of two opposite effects. One the one hand, the sheared granulate is dilated,
but on the other hand the particles reorient and align with respect to the streamlines. Even though aligned cylinders 
in principle may achieve higher packing densities, this alignment compensates for the effect of dilatancy only partially.
The complex rearrangements lead to a depression of the surface above the well oriented region while neigbouring parts still 
show the effect of dilation in the form of heaps.
For grains with isotropic shapes, the surface remains rather flat. Perfect monodisperse spheres crystallize in the
shear zone, whereby positional order partially overcompensates dilatancy effects. However, already slight deviations
from the ideal monodisperse sphere shape inhibit crystallization.}
\vspace{0.5cm}
 \end{@twocolumnfalse}
  ]

%Footnotes
\footnotetext{\textit{$^{a}$Otto-von-Guericke Universit\"at Magdeburg, Institute for Experimental Physics, D-39016 Magdeburg, Germany}}
\footnotetext{\textit{$^{b}$Otto-von-Guericke Universit\"at Magdeburg, Chair for Medical Engineering and Helathcare Telematics, D-39016 Magdeburg, Germany}}
\footnotetext{\textit{$^{c}$Institute for Solid State Physics and Optics, Wigner Research Center for Physics, Hungarian Academy of Sciences, P.O. Box 49, H-1525 Budapest, Hungary }}

%%%%%%%%%%%%%%%%%%%%%%%%%%%%%%%%%%%%%%%%%%%%%%%%%%%%%%%%%%%%%%%%%%%%%%%%%%%%%%%%%%%%%%%%%%%%%%%%%%%%%%%%%%%%%%%%%%%%%%

%---------------------------------------------------------------------------------------------------------
\section {Introduction}

When granular material is exposed to shear flow, it tends to reduce the packing density in the shear zone. This phenomenon is long known, it was first discussed by Reynolds in 1886 \cite{reynolds_1886}. He described the effect of packing density reduction under shear forces. Densely packed granular matter can reduce friction when individual particles are displaced against each other, whereby the occupied volume of the granular bed increases.
Many aspects of this phenomenon have been studied in the past. Bagnold \cite{Bagnold1954,Bagnold1966} reported experiments 
to relate stresses of sheared granulates to the dilation and the shear rate.
Rowe \cite{rowe_1962} performed a systematic study of stress-dilatancy curves for different materials.
Dilatancy of sheared granulates can have dramatic consequences. It influences the strength and the resistance of granular material to slow shear \cite{Dunstan1988}. Its role in the formation of avalanches and the stability of granular heaps \cite{Evesque1991,Rajchenbach1991} and in the stability of geological faults (e. g. Ref.~\cite{Chester1993}) was discussed in literature.
The importance of the role of nonspherical grain shape and its relation to the shear strength and dilatancy of various 
materials escpecially for sand with different particle shapes was also shown 
\cite{rowe_1962,Frossard1979,Bolton1986,Cho2006,Siang2013}.
Experimentally, Thompson and Grest \cite{Thompson1991} established a relation between dilatancy and friction, and they interpreted stick-slip dynamics as a consequence of repeated dilatancy and gravitational compaction phases.
Toiya et al. \cite{toiya_2004} applied high speed imaging of the surface of granular matter in a Taylor-Couette cell
to investigate transient and oscillatory shear flow.
The compaction of the material was derived from the investigation of the height of the granular bed.
The investigated glass beads showed dilatancy effects, and after reversing the rotation direction of the inner cylinder
(i. e. reversal of the shear flow gradient) the beads exhibited compaction and later on expansion again.
Gourlay and Dahle \cite{Gourlay2007} demonstrated that dilatancy is important in high pressure die casting of Al and Mg alloys.
Numerical simulations were performed, e.g. by Bashir et al. \cite{Bashir1991}, who calculated the shearing dilatancy of monodisperse and polydisperse 2D assemblages of disks.
Kruyt and Rothenburg simulated stress-dilatancy relations by DEM, and found an approximately linear relation between dilatancy rate and shear strength \cite{Kruyt2006}.

Different theoretical concepts have been proposed to describe the phenomenon. For example, Coniglio and Herrmann \cite{Coniglio1996} treated the dilatancy in the framework of a phase transition model.
Wan and Guo \cite{wan_1998,wan_1999} developed a model for stress-dilatancy characteristics on the basis of certain constitutive equations.

While earlier experimental studies mainly focused on macroscopic, integral properties like failure of granular beds,
shear forces or surface profiles of the shear zone, modern non-invasive imaging methods allow the study of the
microscopic structure of the granular packing in the shear band. For example, positional ordering in sheared ensembles of
spherical beads has been investigated by an optical fluorescence imaging method
\cite{Tsai2003,Tsai2004,Siavoshi2006,Slotterback2008,Panaitescu2010,Panaitescu2010b,Panaitescu2012,Lorincz2010}.
Nuclear magnetic resonance can be used to reveal the inner structure of a granular bed \cite{Nakagawa1993,Sakaie2008},
even at the grain level  \cite{Finger2006,Fischer2009,Naji2009}.
While Magnetic Resonance Imaging (MRI) relies on the special choice of NMR sensitive grains or an interstitial
NMR sensitive liquid, X-ray computed tomography (CT) studies \cite{Zhang2006,Kabla2009,Borzsonyi2012,Wegner2012}
offer the advantage of a much broader choice of suitable materials.

A quasi-2D system was studied experimentally by O'Sullivan et al.~\cite{OSullivan2002}, using coaxial (initially hexagonally packed) rods that were sheared normal to the axis direction. In that geometry, it is possible to observe the rod packing optically from a side view of the shear cell.

Using refractive index matching and an optical tomographic technique, the positions of individual grains could be resolved
inside of the material \cite{Tsai2003,Tsai2004}, and it was shown that sheared granular materials form homogeneous
crystalline nuclei. It passes through an inhomogeneous nucleation at the container side walls.
A different study showed that vibration can also induce crystallization of monodisperse spheres,
which can be melted by sufficiently strong shear \cite{Daniels2005,Daniels2006}.

Linear shear cells offer a simple geometry for experiments, but in practice only a limited displacement can be achieved. Some experiments that were focused on shear induced order were performed under cyclic shear in a special geometry \cite{Nicolas2000,Panaitescu2010b,Panaitescu2012}.  If one is interested in the structure of the asymptotic state, it is more suitable to use a cylindrical geometry. Annular Couette geometries \cite{Bagnold1954,Tsai2003,Tsai2004,Daniels2005,Daniels2006} have the advantage that they are not restricted in achievable strain.
During the last decade, cylindrical split-bottom containers
\cite{Fenistein2003,Fenistein2004,Unger2004,Unger2007,Sakaie2008,Cheng_2006, Fenistein2006,Dijksman2010} have been proven
to provide a convenient geometry where practically unlimited displacements can be achieved between an inner,
rotating granular volume and an outer, fixed volume. The geometry and width of the shear zone can be controlled to a
certain extent by the ratio of the filling height $h$ and the radius $r$ of the rotating bottom disk
\cite{Unger2004, Unger2007,Cheng_2006,Fenistein2004,Fenistein2006,Sakaie2008}. If this ratio is smaller than 0.6,
the central core rotates with the bottom disk, and the sheared and dilated zones essentially coincide \cite{Sakaie2008}.
For higher granular beds ($h/r \ge 0.75$), there is a relatively long initial transient after which the zone reaches its stationary state, therefore the resulting dilated zone deviates substantially from the late-time shear zone \cite{Sakaie2008}.

The effect of dilatancy was studied in details by Sakaie et al. \cite{Sakaie2008} with poppy seeds of diameter of about 1 mm by means of MRI in a split bottom shear cell. The spatial structure of the dilated zone and its evolution with applied strain were investigated. While the experimental resolution was not sufficient to separate the individual grains, the NMR signal provided reliable quantitative packing density data. The main results of this study are that (1) the dilated zone spreads rather far through the system and (2) the amount of dilatancy initially grows with accumulated strain, until it saturates gradually for strain of order one.

These previous investigations dealt with spherical or nearly spherical grains. In our present study, we compare the packing of elongated (almost cylindrical), oblate and spherical particles under shear in a split bottom container. We investigate the samples by means of X-ray CT to record the individual particle positions and orientations, to derive their packing densities in the shear zone, and to extract the height profile of the granular bed. The individual particles have dimensions of the order of several millimeters, they can be well distinguished within the available spatial resolutions of the CT scanner. The experiment provides the options both to study the packing on the level of individual particles, and to average over particular zones of the sample and/or sequences of tomograms.

With non-spherical particles, we expect qualitatively new effects evoked by the induced orientational order in the shear
zone \cite{Borzsonyi2012,Borzsonyi2012a,Wegner2012,Borzsonyi2013}. The two competing processes are the dilatancy that tends to reduce the packing density in the shear zone, and the induced orientational order of anisometric grains that allows denser packing. In this study, we analyze cylinders with different aspect ratios, oblate ellipsoids, and spherical particles. Two types of sphere-shaped grains are compared, one sample consists of perfectly sphere-shaped monodisperse grains (airsoft balls) and another one of non-perfect shape, with a small polydispersity (peas).
By comparison of these samples, we try to separate grain shape influences on dilatancy, on orientational and positional order.

%---------------------------------------------------------------------------------------------------------
\section {Experiment}

\begin{figure}[htbp]
	\includegraphics[width=0.92\columnwidth]{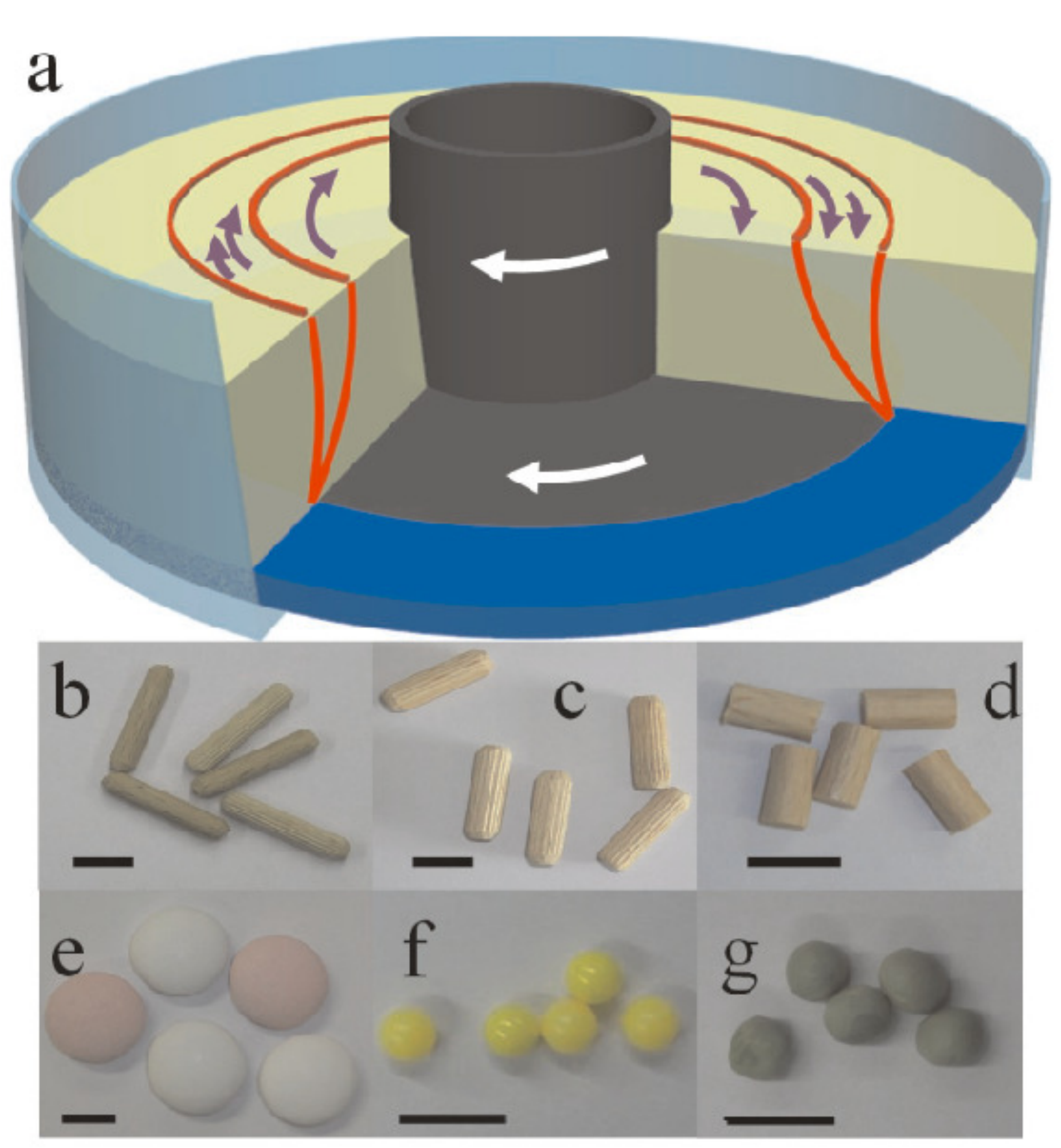}
	\caption{(a) Schematic drawing of the split bottom shear cell, dark gray (online: red) curves in the bulk indicate 
position of the shear zone. (b-g) Photographs of particles studied, pegs with aspect ratios  $Q=\ell/d$  
of (b) 5, (c) 3.3 and (d) 2, (e) lentils, (f) spheres and (g) peas. The horizontal bars correspond to 1~cm.}
	\label{fig:aufbau2}
\end{figure}

The experiments were performed with a cylindrical split bottom shear container of radius of 28.5~cm,
see sketch in Fig.~\ref{fig:aufbau2}a.
The circular plate (radius 19.5 cm) at the bottom of the container and the inner cylinder (radius 8 cm) were rotated
together manually during the experiments. The shear container is filled with the granular material up to a
uniform height between 5 cm and 6.5 cm, depending on the experiment.
This corresponds to a relatively low filling ratio of $h/r \le 0.33$.
Even though we take particular care during the filling procedure to achieve orientationally disordered packings,
a slightly biased ordering of the elongated and flattened grains cannot be completely avoided.
This is not critical for the shear experiments since the shear realigns the grains in the regions of interest,
irrespective of the pre-existing orientation.

The materials used in this study are shown in Fig.~\ref{fig:aufbau2}b-g. To characterize the deviation from the 
spherical shape we use two quantities (i) the aspect ratio $Q$, which is the ratio of the size of the particle along
its rotational axis and perpendicular to it and (ii) the sphericity $\Psi$ which is the ratio of the surface area 
of the particle and a sphere with the same volume. The list of the materials used is as follows:
%The materials used in this study are:
\begin{itemize}
\item
wooden pegs with cylinder shape and tapered ends, diameter $d=5$~mm, length $\ell=25$~mm, aspect ratio $Q=\ell/d= 5$, 
sphericity $\Psi= 0.726$
\item
wooden pegs with cylinder shape and tapered ends, diameter $d=6$~mm, length $\ell=20$~mm, $Q\approx 3.3$,
$\Psi= 0.803$
\item
wooden pegs with cylinder shape,  diameter $d=5$~mm, length $\ell=10$~mm,  $Q=2$,
$\Psi= 0.832$
\item
oblate ellipsoidal chocolate lentils ({\em Piasten}) covered with hard icing with 
diameter $d=18.5$~mm, axial length $h=8.3$~mm, aspect ratio 
$Q=h/d=0.45$, and $\Psi= 0.886$
\item
airsoft munitions, perfect spheres, monodisperse, $d=6$~mm
\item
peas, with small deviations from a perfect sphere, polydisperse with mean diameter of 7.6~mm, standard deviation 0.23~mm,
$\Psi= 0.998$
\end{itemize}
The intermediate ($Q=3.3$) pegs and some of the long ($Q=5$) and the short ($Q=2$) pegs have slight axial groves on 
their surfaces.

After filling the particles in the container, the bottom disk and inner cylinder are rotated manually as indicated by white arrows in Fig.~\ref{fig:aufbau2}. The rotation rate was not recorded, but was kept well within the quasi-static regime.
By rotating the inner part of the setup, a more or less vertical zone with radial shear gradient forms in the
material, where the strain is localized.
The location of this zone is indicated (red) in Fig.~\ref{fig:aufbau2}.
At regular intervals, the rotation is stopped to record an X-ray tomogram of a representative section of the shear cell.
We use the robot-based flat panel X-ray C-arm system Siemens Artis zeego of the INKA lab, Otto von Guericke University, Magdeburg. The spacial resolution can be chosen as 2.03 pixel/mm or 1.48 pixel/mm, which corresponds to recorded volumes of 25.2 cm $\times$ 25.2 cm $\times$ 19 cm or 34.8 cm $\times$ 34.8 cm $\times$ 24.4 cm, respectively.
The 3D arrangement of the particles is then determined experimentally from the X-ray tomogram.

Measurements during the initial transients towards the stationary asymptotic state are single scans, their statistics
is comparably poor. The measurements in the asymptotic state were statistically averaged over at least  125  tomograms for each grain type. By this averaging, we obtain a good representation of the average order, alignment and packing density in the shear zone and in the rotated inner part. Between the shear zone and the outer container wall, there is only little motion, and we essentially observe the initial random particle distribution. Some details will be discussed below.

In order to calculate the packing density in different parts of the shear cell, we binarize the tomograms.
We define a threshold for all voxels, that determines whether the voxel belongs to a grain or not.
The correct choice of this threshold is essential for the absolute packing densities determined from the tomograms.
Our algorithm is based on Otsu's method \cite{Otsu1979} and special care is taken to only take into account the
useful part of the tomogram, i.e. the binarization threshold is determined only from the particles and interstitial gas,
while border regions (container wall, rotating plate) are avoided.
Afterwards the binarized tomograms are averaged for each material. We are interested in the packing densities
in given positions in the shear zone.
Therefore, data in each 3D image are projected to a 2D representation where
we average over all equivalent voxels, i. e. voxels with the same distance to the center of rotation and the same height.
This method gives an accurate measure of the relative packing densities in different parts of the shear zone.

%%% FIG 2 %%%%%%%%%%%%%%%%%%%%%%%%%%%%%%%%%%%%%%%%%%%
%
\begin{figure}[htbp]
	\centerline{\includegraphics[width=0.9\columnwidth]{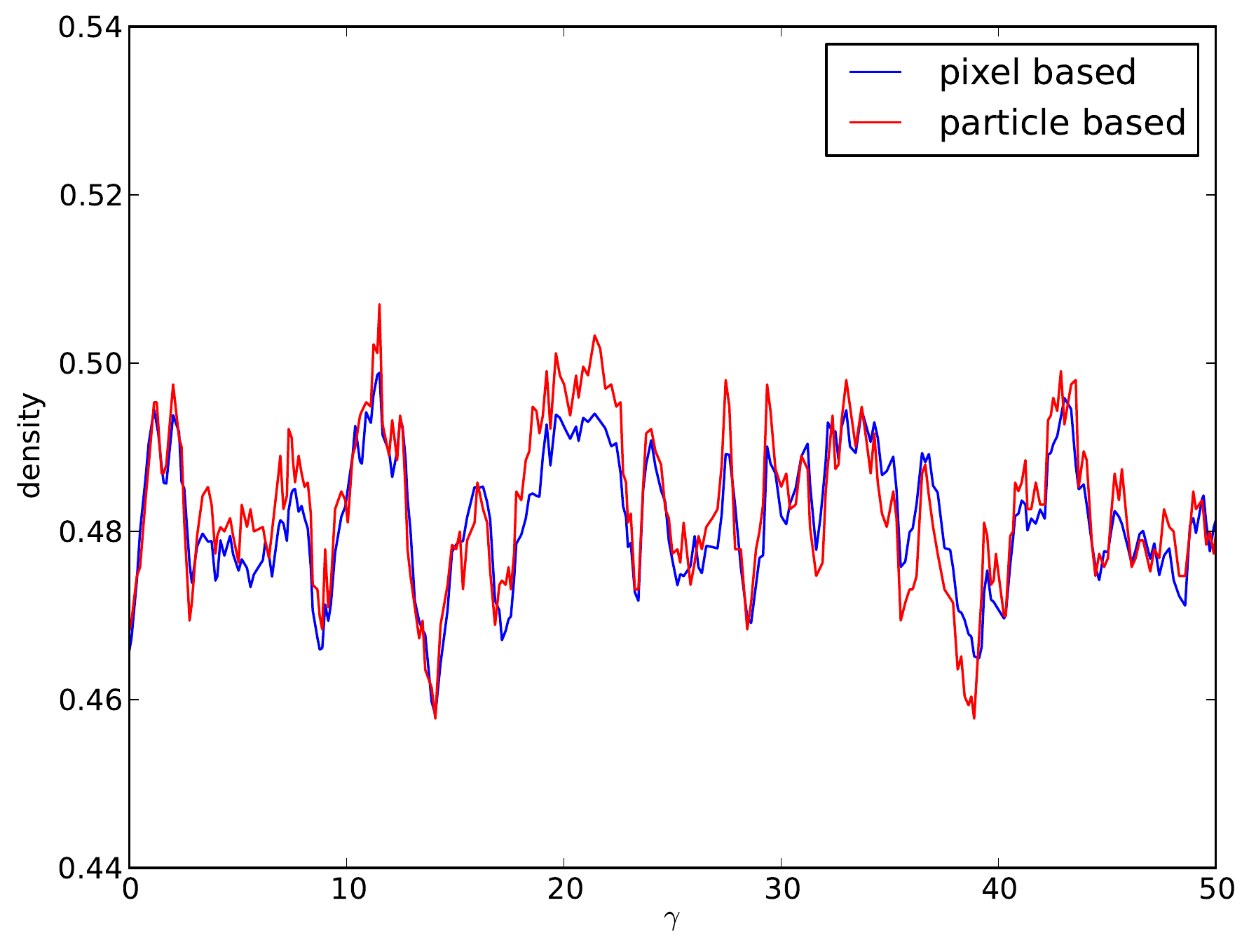}}
	\caption{Packing density in the shear zone as a function of the maximal local strain $\gamma$ for pegs with $Q=5$. The two 
curves correspond to the two methods:  density obtained from (i) binarized images (Otsu's method) and (ii) particle 
identification.}
	\label{fig:twomethods}
\end{figure}
The accuracy of our method can be demonstrated by comparing the result to density data obtained by a different procedure
where the particles are identified. Our particle tracking code determines the center of mass and the orientation of each 
individual particle. The density of the sample in a given volume can be calculated using this information. Fig. \ref{fig:twomethods} 
shows the density of the shear zone for pegs with $Q=5$ as a function of the maximal local strain $\gamma$, as the sample 
is sheared. 
Here the value of $\gamma$ has been determined from the particle displacements (as described in 
\cite{Borzsonyi2012,Borzsonyi2012a}).
As it is seen, the density of the zone obtained by these two methods is very similar with noticeable fluctuations.
The corresponding volume contained about 180 particles. In this paper we are mostly interested in the time averaged
density under stationary shear which allows us to reduce the noise (arising due to fluctuations) by averaging subsequent 
tomograms. 
The advantage of the the first method - direct determination of the packing density from the binarized tomograms - over particle 
identification is its higher efficiency at comparable precision.
Thus in the rest of this paper we determine the density directly 
from the binarized tomograms.

%---------------------------------------------------------------------------------------------------------
\section {Results and Discussion}

%%% FIG 3 %%%%%%%%%%%%%%%%%%%%%%%%%%%%%%%%%%%%%%%%%%%

\subsection{Individual Particle Arrangement}
\begin{figure*}[htbp]
	\centerline{\includegraphics[width=0.8\textwidth]{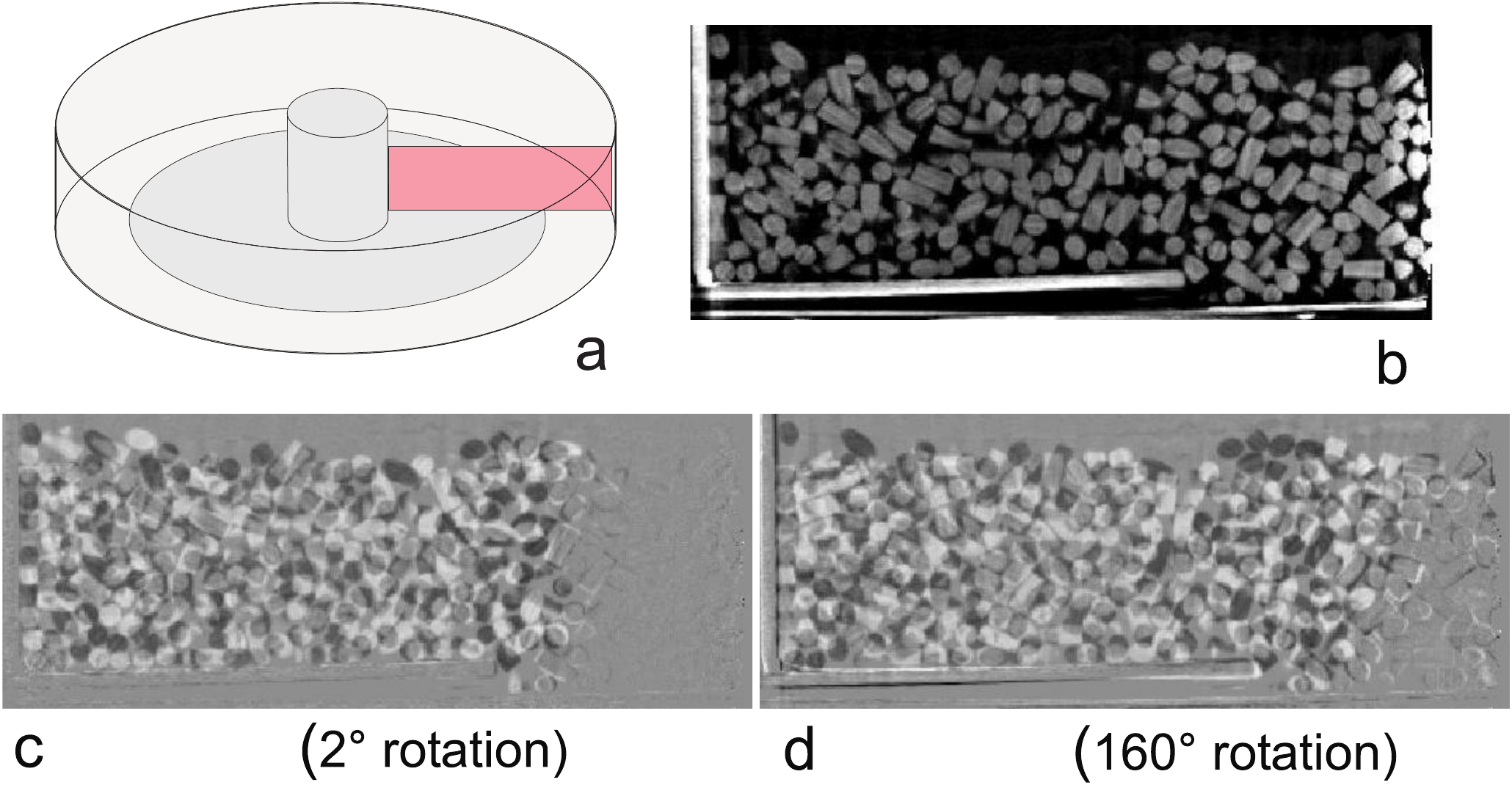}}
	\caption{Radial cut through the tomogram along a vertical slice for pegs with aspect ratio $Q=2$. 
Image (b) was taken after shearing the sample for about 20$^\circ$, images (c) and (d) show differences between that slice and a slice recorded after a 2$^\circ$ rotation, and 160$^\circ$ rotation, respectively. Original positions appear darker, new positions brighter. Image sizes are 19 cm $\times$ 7.5 cm, the rotating plate is seen at the bottom. }
	\label{fig:radiale}
% Bilder 530 x 210 Pixel, Skalenfaktor 0.0357 cm/Pixel
\end{figure*}

Figure \ref{fig:radiale} shows an exemplary radial cut (i.e. a vertical plane normal to the flow) of the tomograms.
The material is pegs with length 1 cm and aspect ratio $Q=2$.
When the inner cylinder is rotated, the material in the center of the sample rotates as a solid body, the material
near the outer border is not moving, while the section in between is sheared.
Thus the material in the shear zone and within the central area is moved, the grains outside remain essentially fixed, which is
nicely visualized on the two difference images Fig.~\ref{fig:radiale}c and Fig.~\ref{fig:radiale}d.
The first difference image \ref{fig:radiale}c was obtained after a relatively small rotation, here one can identify the outer border of the shear zone.
It is also seen, in Fig.~\ref{fig:radiale}d which shows the changes after almost a half turn of the inner plate, that there are gradual reorientations of the particles (creeping motions) also outside the kernel of the shear zone. These gradual displacements can modify the packing density even far outside the region where shear is focused. When we refer to the shear zone in the following, we use a practical definition: particles that are displaced by more than one particle diameter with respect to their neighbors form the shear zone.

\subsection{Packing Density, Alignment and Crystallization}

\begin{figure*}[htbp]
	\includegraphics[width=\textwidth]{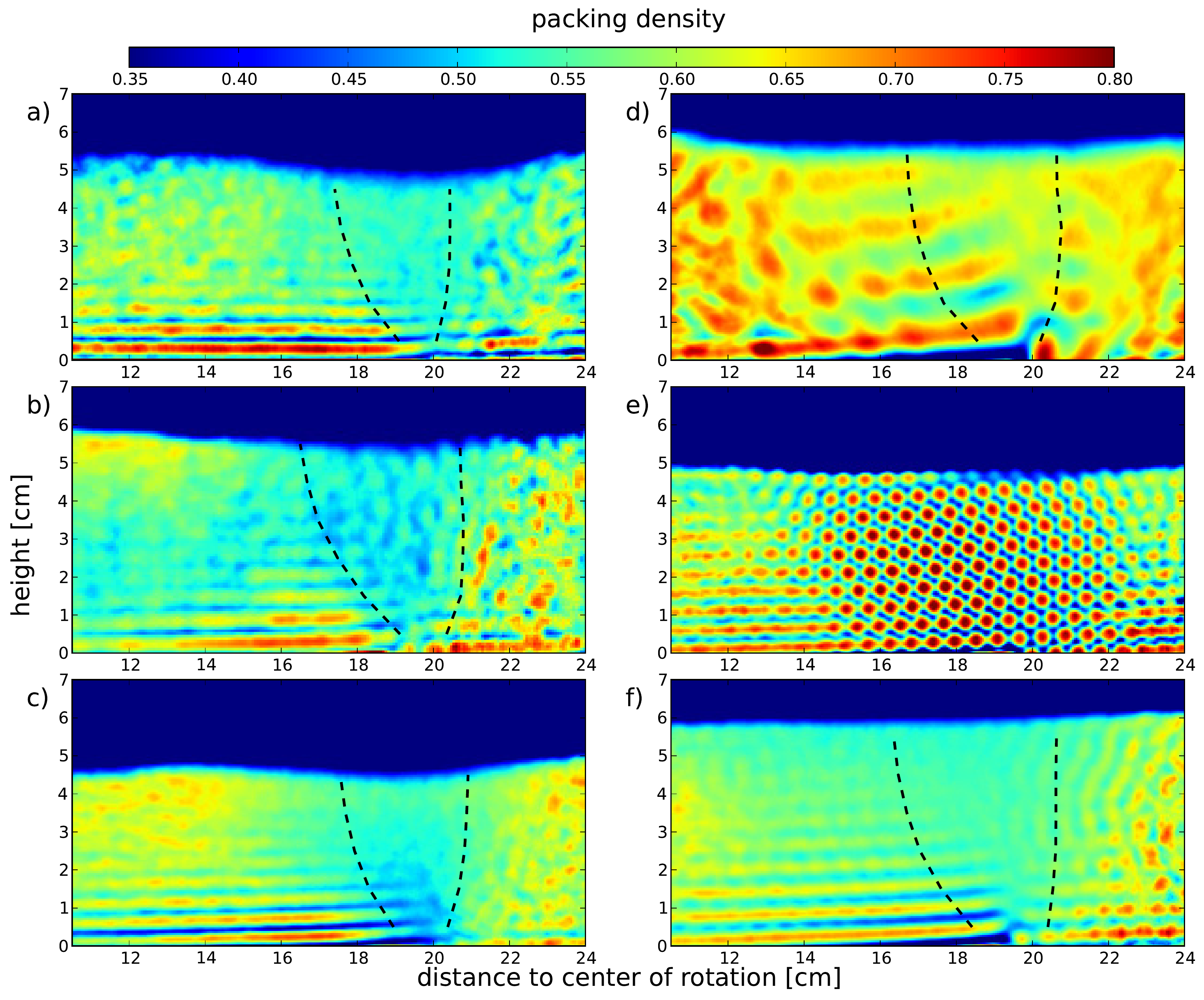}
\centerline{\includegraphics[width=0.36\textwidth]{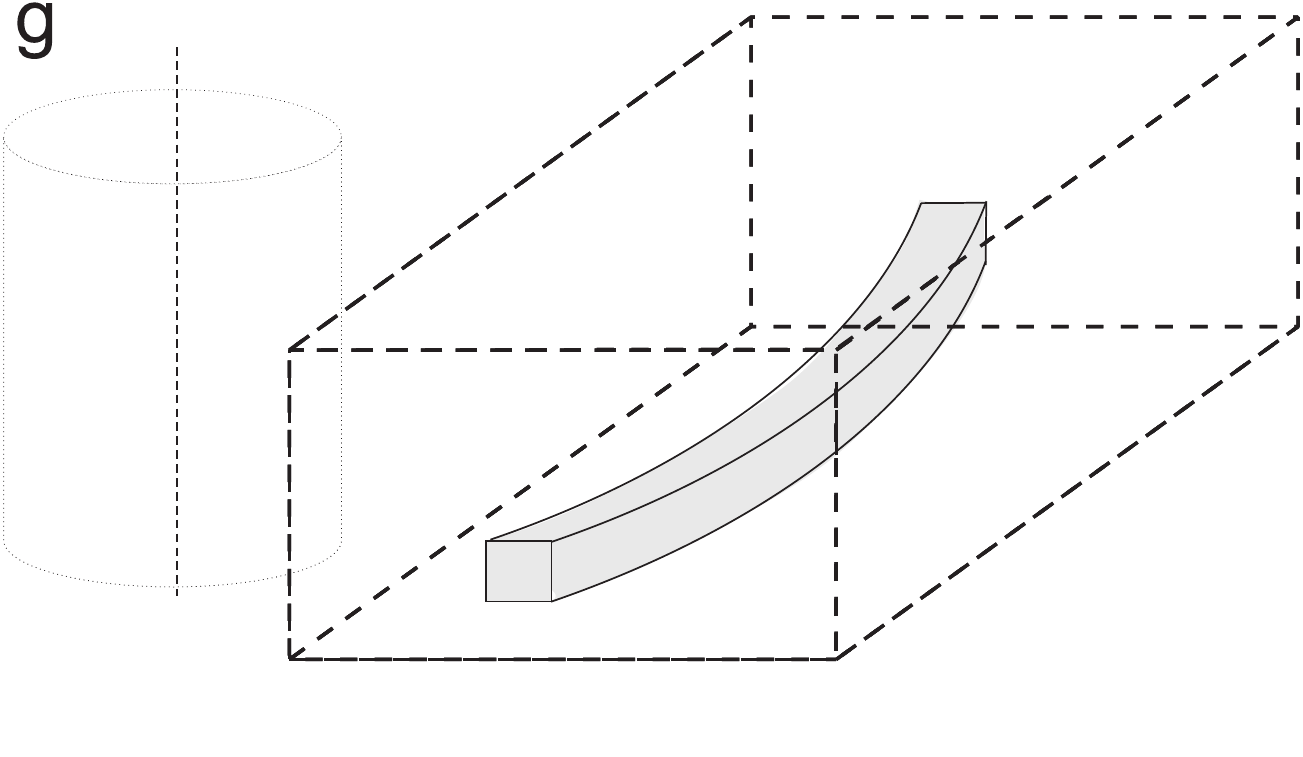}}
\caption{Packing density for elongated particles with aspect ratio a) $Q=5$,
b) $Q=3.3$ and c) $Q=2$, d) lentils, e) spheres, and f) peas in a cylindrical
shear cell, the split bottom edge is located at 19.5 cm. The sketch g)
visualizes the shape of the zones that are averaged for each pixel in the
images, the dashes mark the tomogram dimensions.
	\label{fig:dichten_alle}}
\end{figure*}

The main result of this work can be seen in Figure \ref{fig:dichten_alle}, which depicts the averaged local
packing density (i.e. the probability to find a particle) in a given height and distance from the
central axis of the container. In order to obtain these images, a large number of tomograms ($>125$) for
each grain type have been evaluated, the average is over regions with equal distance from the rotation
center (see sketch in Fig~\ref{fig:dichten_alle}g). There are several interesting aspects in these images,
the spatially resolved packing densities and orientational ordering in the shear zone, as well as
the height profiles of the granular bed.

%%% FIG 4 %%%%%%%%%%%%%%%%%%%%%%%%%%%%%%%%%%%%%%%%%%%

The first interesting aspect in Fig.~\ref{fig:dichten_alle} is, that the packing density in the shear zone is clearly
reduced both for anisometric particles with different aspect ratios (Fig.~\ref{fig:dichten_alle}a-d) and for the nearly
spherical particles (Fig.~\ref{fig:dichten_alle}f).
There are two competing influences, related to the reduction of friction by dilatancy and the reduction of friction by
alignment. While dilatancy is related to decreasing packing density, the alignment allows even denser packing.
But even though the elongated and flattened grains show a pronounced orientational order, which has been quantitatively
determined in previous work \cite{Borzsonyi2012,Borzsonyi2012a,Wegner2012}, the average packing density is found to be
lower than in the non-sheared, non-aligned material. This will be discussed in more detail below.
The effect is independent of whether the grains are oblate or prolate. The flattened lentils also align in the shear
field, but they also reduce their packing density by a comparable amount.

Note that we can only compare relative packing densities on an exact quantitative level, the absolute packing densities may be affected by some systematic errors: First, the initial packing densities are difficult to control in the container,
deviations from random orientations arising from the filling procedure (see, e.g. \cite{Buchalter1992,Buchalter1994}) are preserved in the non-sheared parts of the sample. Second, one has to consider a small systematic error related to the difficulties in the exact choice of the threshold parameters in the binarization of images, mentioned in the previous section. Moreover, the packing is affected by the boundary at the bottom plate. This is particularly evident at the left hand sides of the images in Fig. \ref{fig:dichten_alle}, 'inside' the rotating region. There, each tomogram captures different particles and thus the surface-induced layering is particularly evident in the averaged images.
It appears as a density modulation in vertical direction. For the cylinders, the wavelength of this modulation deviates less than 5~\% from the diameter, it is
$(5.1\pm 0.2)$~mm for the $Q=5$ pegs,
$(5.9\pm 0.2)$~ mm for the $Q=3.3$ pegs,
$(4.9\pm 0.2)$~ mm for the $Q=2$ pegs). The layer widths for the perfect spheres (airsoft balls), $(5\pm 0.2)$~mm, and for the peas, $(6\pm 0.2)$~mm, are about 20~\% lower than the respective grain diameters.

The observed packing density modulation with a characteristic length of less than one particle diameter
is expected for the spherical particles, since the second layer may partially penetrate the first one.
For the cylinders, however, it is reasonable that the smallest diameter is relevant, since the particles arrange preferentially horizontally after filling. For the more or less azimuthally disordered cylinders, the period of the vertical density modulation is expected to be
close to the  diameter of the particle.
For the lentils, the wavelength of the packing density modulation strongly increases when going from the non-sheared region
towards the shear zone. This is a consequence of the fact, that the particles in the non-sheared regions are more or less
oriented with their short axes vertically as a result of the initial filling procedure, while in the shear zone, their short axis is preferentially aligned close to the shear gradient \cite{unpub} (i.e. more or less horizontal).

%%% FIG 5 %%%%%%%%%%%%%%%%%%%%%%%%%%%%%%%%%%%%%%%%%%%
\begin{figure}[htbp]
\centerline{	\includegraphics[width=0.8\columnwidth]{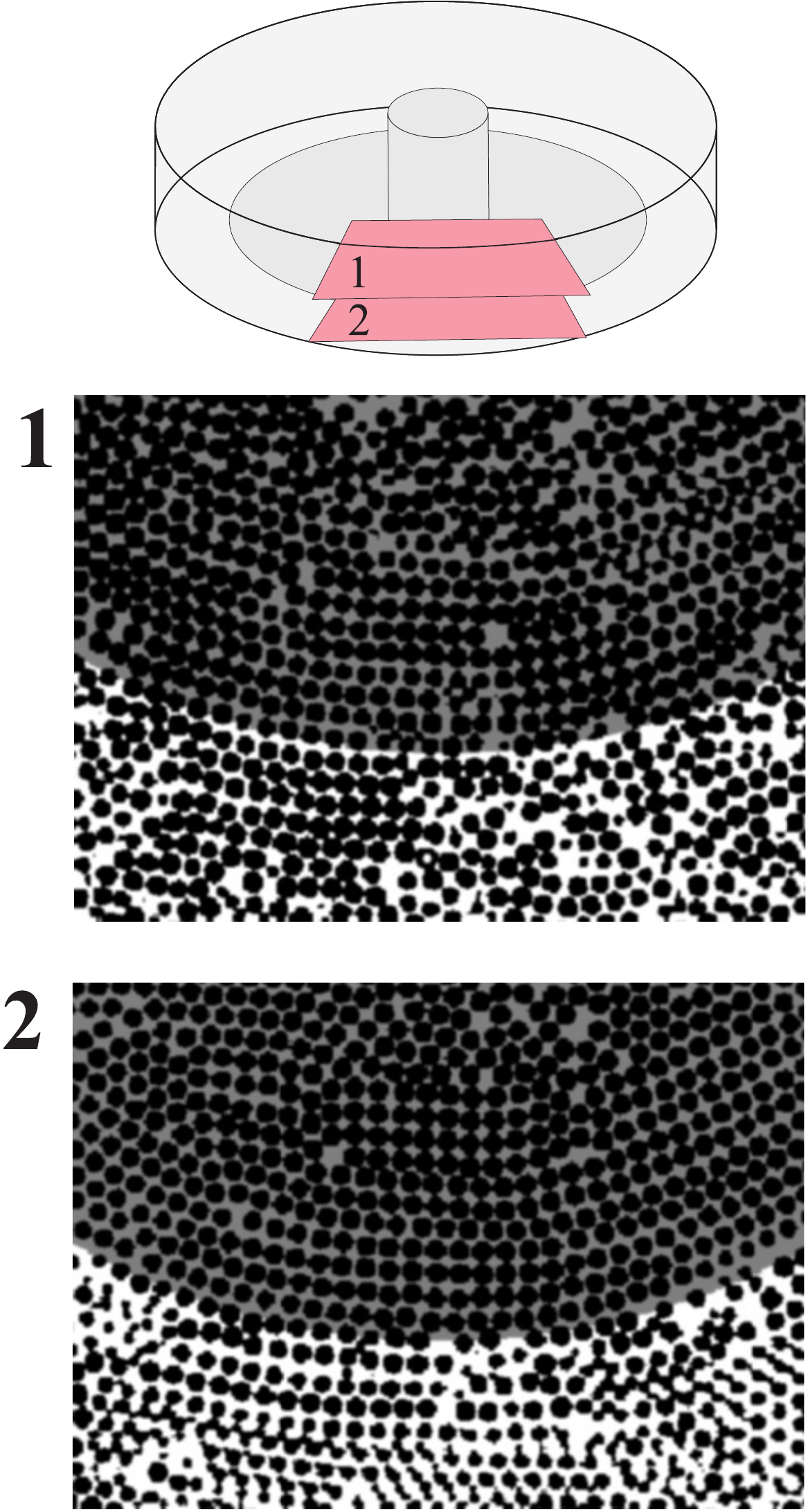}}
	\caption{Packing of perfect spheres (airsoft balls) in the asymptotic state. The figure shows two horizontal slices taken from one individual tomographic picture, at the middle height of the granular bed (1) and at one particle radius above the bottom disk (2). The grey shade visualizes the rotating bottom plate. The top sketch indicates the positions of the two slices.
	\label{fig:ellak1}}
\end{figure}
\begin{figure}[htbp]
\centerline{	\includegraphics[width=0.99\columnwidth]{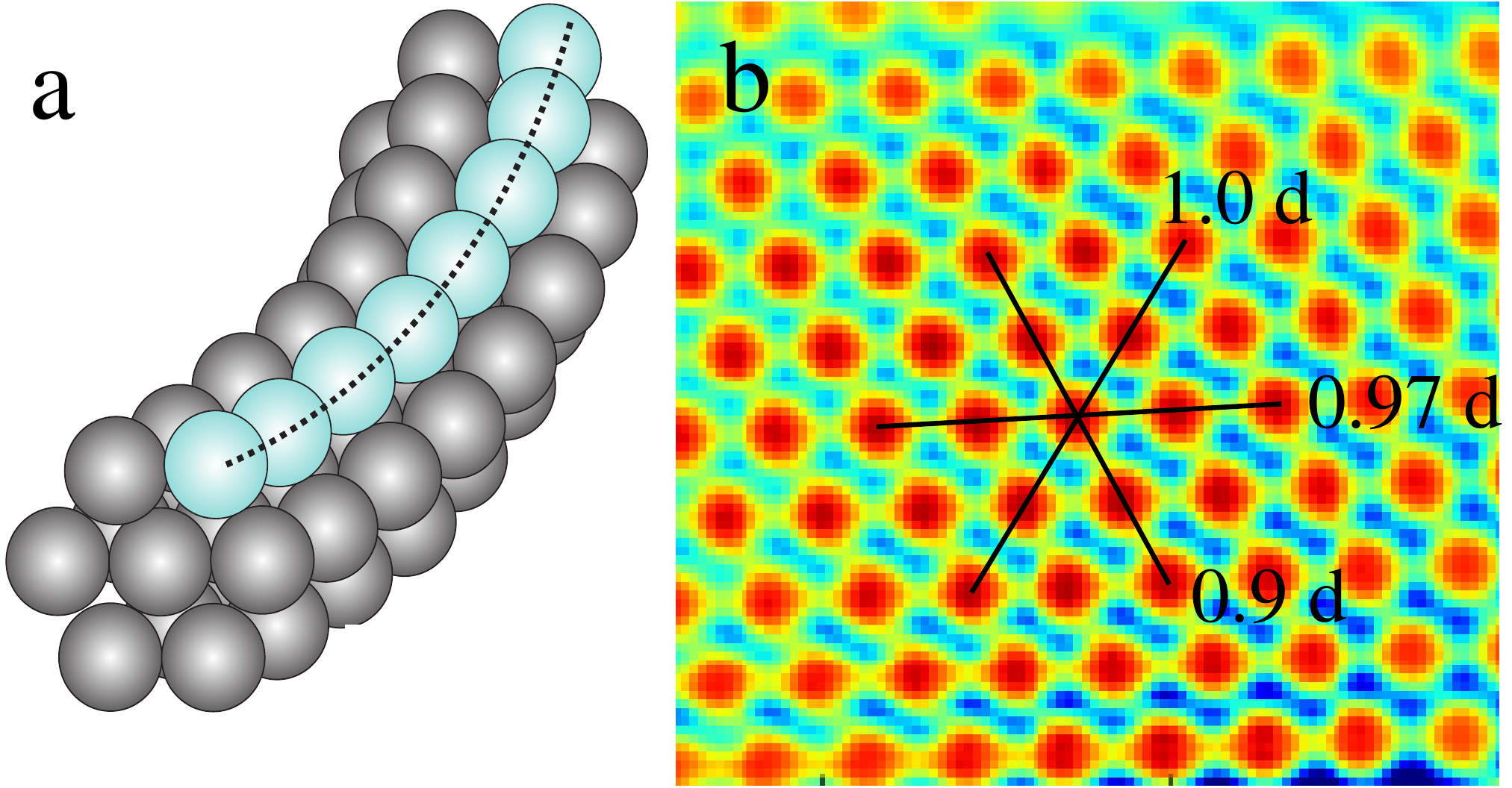}}
	\caption{(a) Schematic sketch of the arrangement of the perfect spheres (airsoft balls) in the shear zone. 
The dashed line indicates the tangential direction of shear flow.
(b) Packing of perfect monodisperse spheres (6 mm {airsoft} balls) in the shear zone,
magnified detail of Fig.~\ref{fig:dichten_alle}e. The numbers show the distances to the nearest neighbors along the
marked axes, in units of the grain diameter. Color coding as in Fig.~\ref{fig:dichten_alle}.
	\label{fig:hexagons-lattice}}
\end{figure}

A second interesting aspect in Fig.~\ref{fig:dichten_alle} is the three-dimensional positional ordering of the perfect
spheres in the shear zone ( Fig.~\ref{fig:dichten_alle}e). There are several characteristic features of this
arrangement that deserve discussion.
The first thing to note is, that there is no obvious average density decrease in the zone (unlike the other samples),
we will come back to this question in Sec. \ref{Dilation}.
The periodicity of the structure is nearly equal to the sphere diameter (see details below). The spheres form a well-defined
hexagonal lattice in the planes normal to the flow direction. A cut along horizontal planes inside the bulk, normal to the 
neutral direction (Fig.~\ref{fig:ellak1}), evidences that the spheres chain up in tangential direction. The chains follow 
the shear flow direction. Thus, the complete 3D structure is that of bent hexagonally packed chains of spheres along 
the tangential flow direction, as sketched in Fig.~\ref{fig:hexagons-lattice}a.
In this structure, the chains of spheres can slide along each other, thus friction is reduced with respect to a 
disordered packing. Similar observations have been reported in optical investigations with \cite{Daniels2005,Daniels2006}
and without \cite{Tsai2003,Tsai2004} additional vibrations.
We note that this lattice is not only present in the core of the shear zone but it extends well into the adjacent regions 
where the grains are not sheared (i.e. they are displaced respective to their neighbors by much less than one particle 
diameter).

%%% FIG 6 %%%%%%%%%%%%%%%%%%%%%%%%%%%%%%%%%%%%%%%%%%%

A detailed quantitative inspection of the lattice parameters reveals that the hexagonal lattice is not exactly equilateral.
Figure \ref{fig:hexagons-lattice}b shows a magnified detail of the center of the shear zone in Fig.~\ref{fig:dichten_alle}e,
with three lattice directions and the corresponding distances of neighboring chains. It reveals that some chains 
partially overlap: in one of the directions the distance is significantly smaller than the sphere diameters.
In order to reduce friction in a shear plane, it suffices that two lattice constants are comparable to the grain diameter.
The direction with the shortest chain distance of 0.9 $d$, in which the sphere chains partially interdigitate,
lies roughly perpendicular to the shear gradient (cf. Fig.~\ref{fig:dichten_alle}).

\begin{figure}[htbp]
\centerline{	\includegraphics[width=0.8\columnwidth]{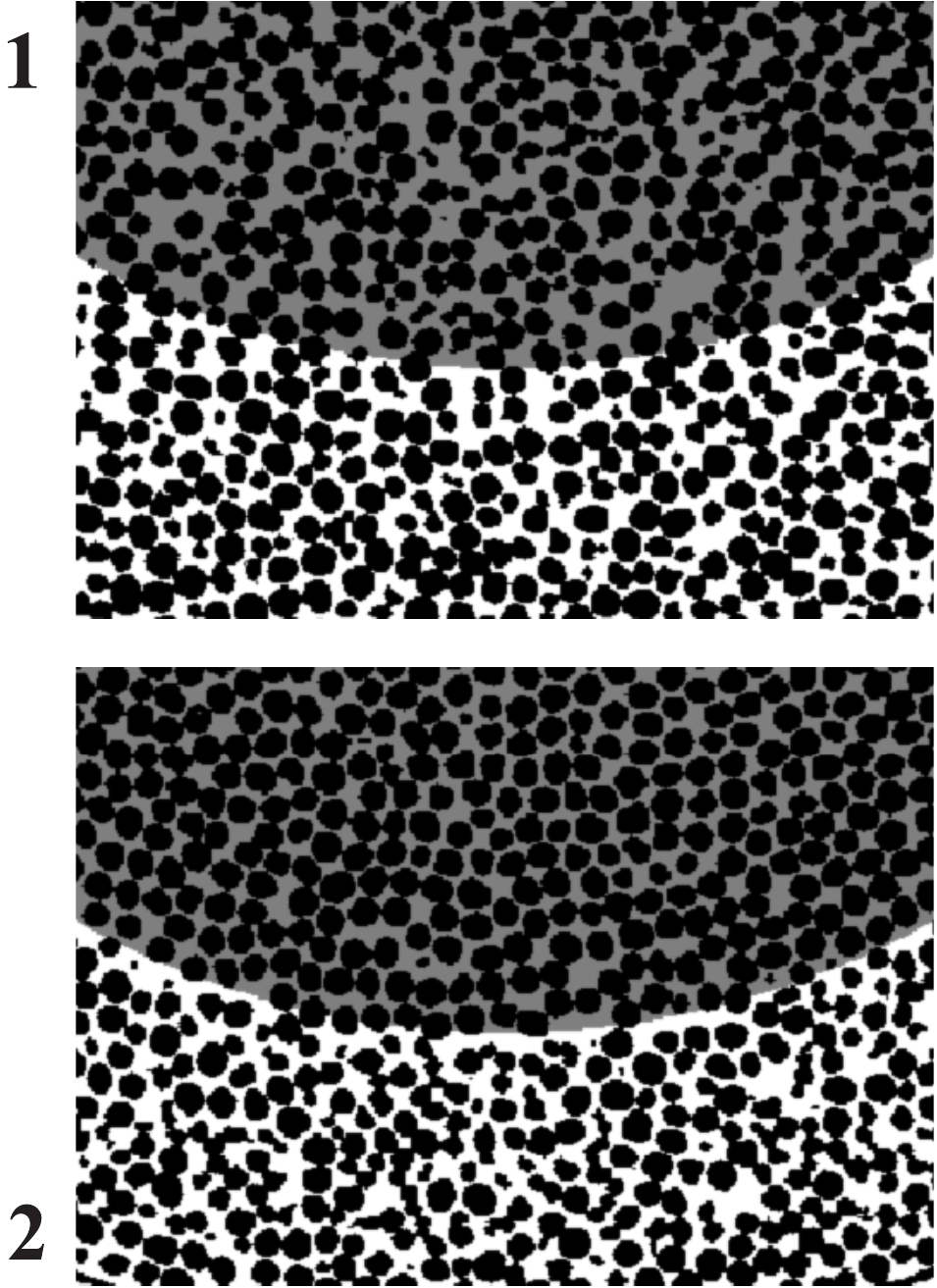}}
	\caption{Packing of peas in the asymptotic state, horizontal slice taken from a single tomogram
at the middle height of the granular bed (1) and at one particle radius above the bottom disk (2), see sketch in Fig.~\ref{fig:ellak1} for the positions of the slices.
	\label{fig:ellak2}}
\end{figure}

Even a slight deviation from monodispersity and/or from perfect sphere shapes (standard deviation $\pm 3$~\% for the radii of the peas) destroys this regular structure completely. Figure \ref{fig:ellak2} shows the same slices as in Fig.~\ref{fig:ellak1}, taken from a tomogram of sheared peas. A slight ordering is indicated only in the first layer above the bottom plate, but inside the bulk, positional order is practically absent.

%%% FIG 7 %%%%%%%%%%%%%%%%%%%%%%%%%%%%%%%%%%%%%%%%%%%

\begin{figure*}[htbp]
\centerline{	
\includegraphics[height=0.32\textwidth]{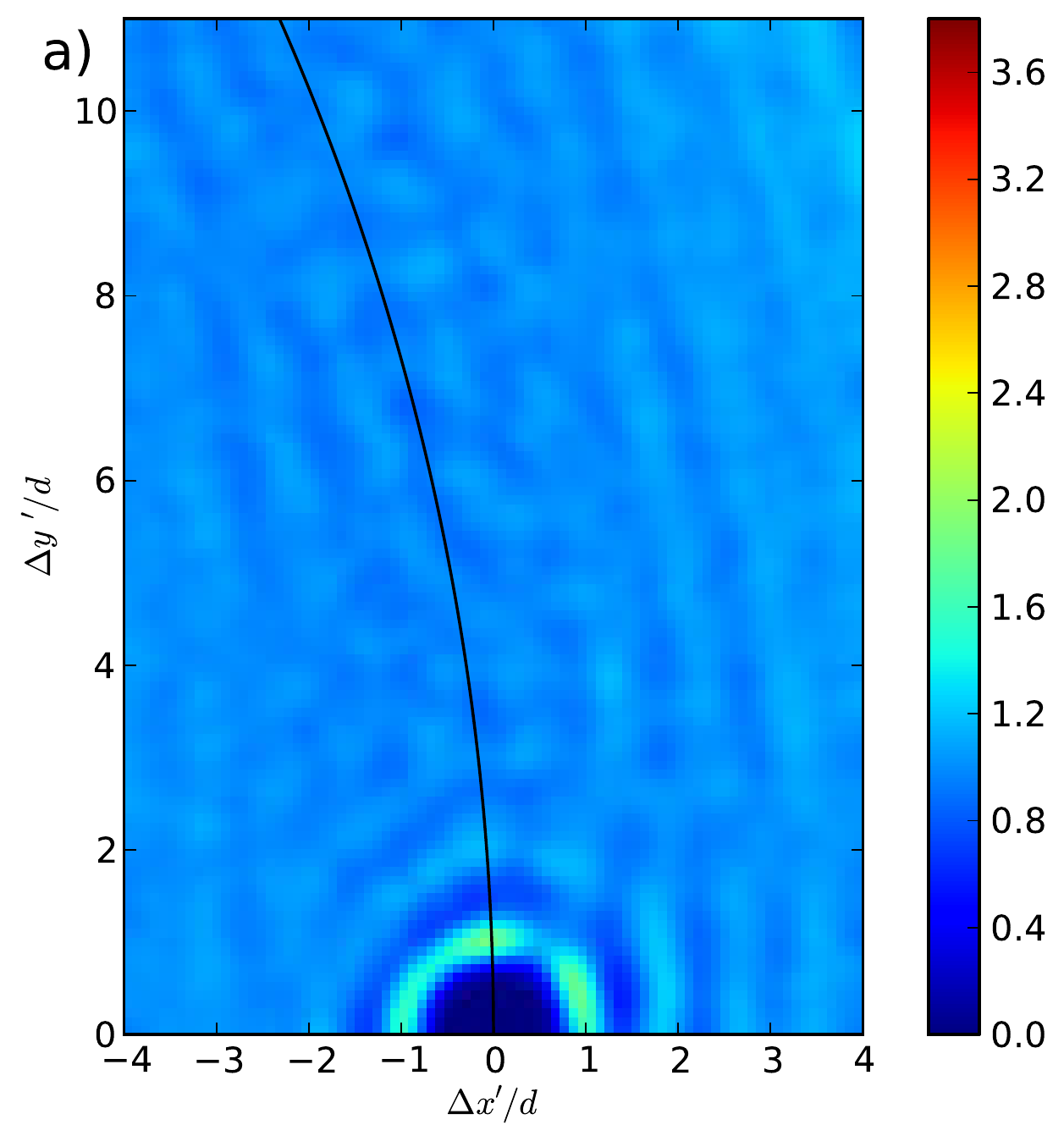}
\hfill
\includegraphics[height=0.32\textwidth]{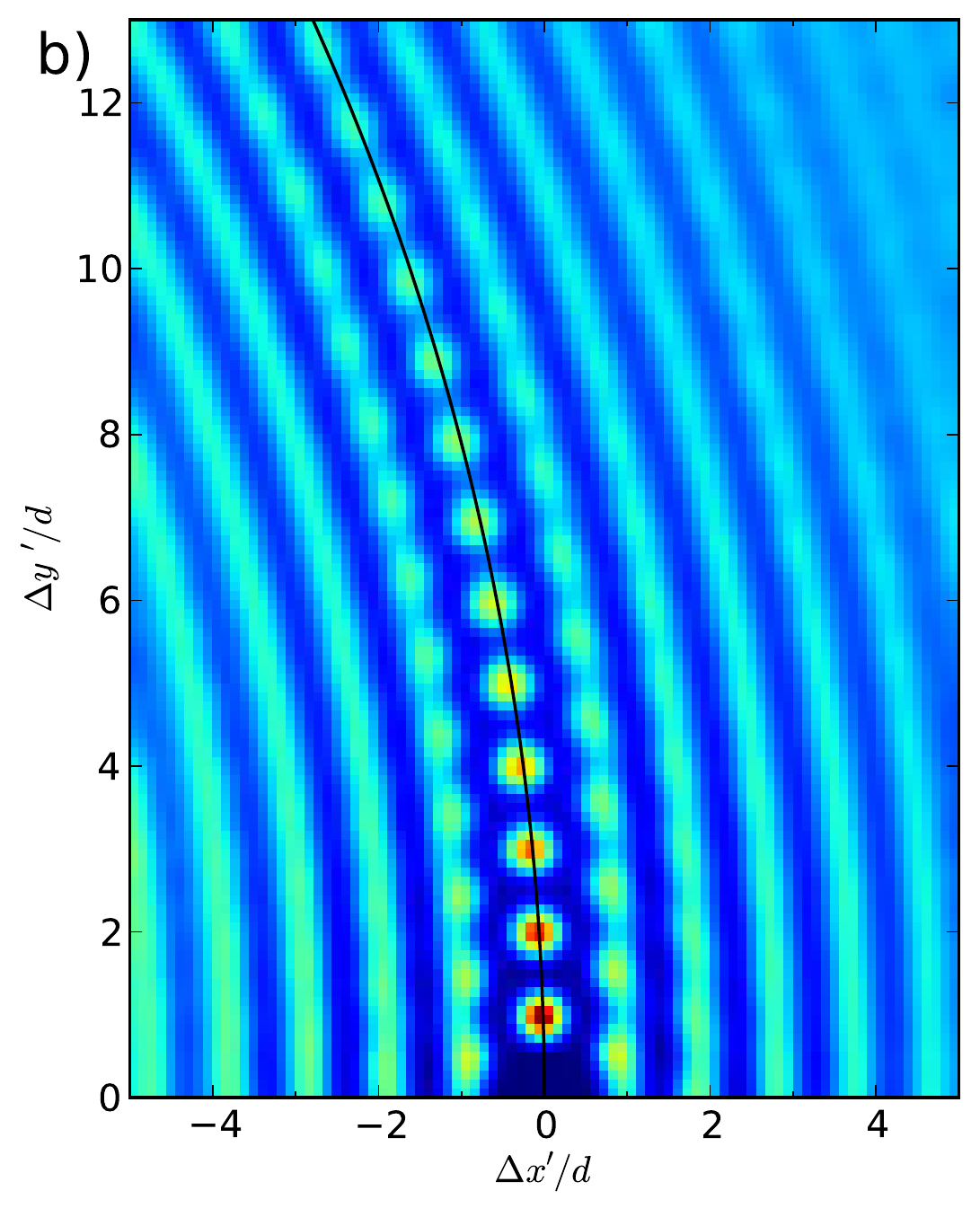}
\hfill
\includegraphics[height=0.32\textwidth]{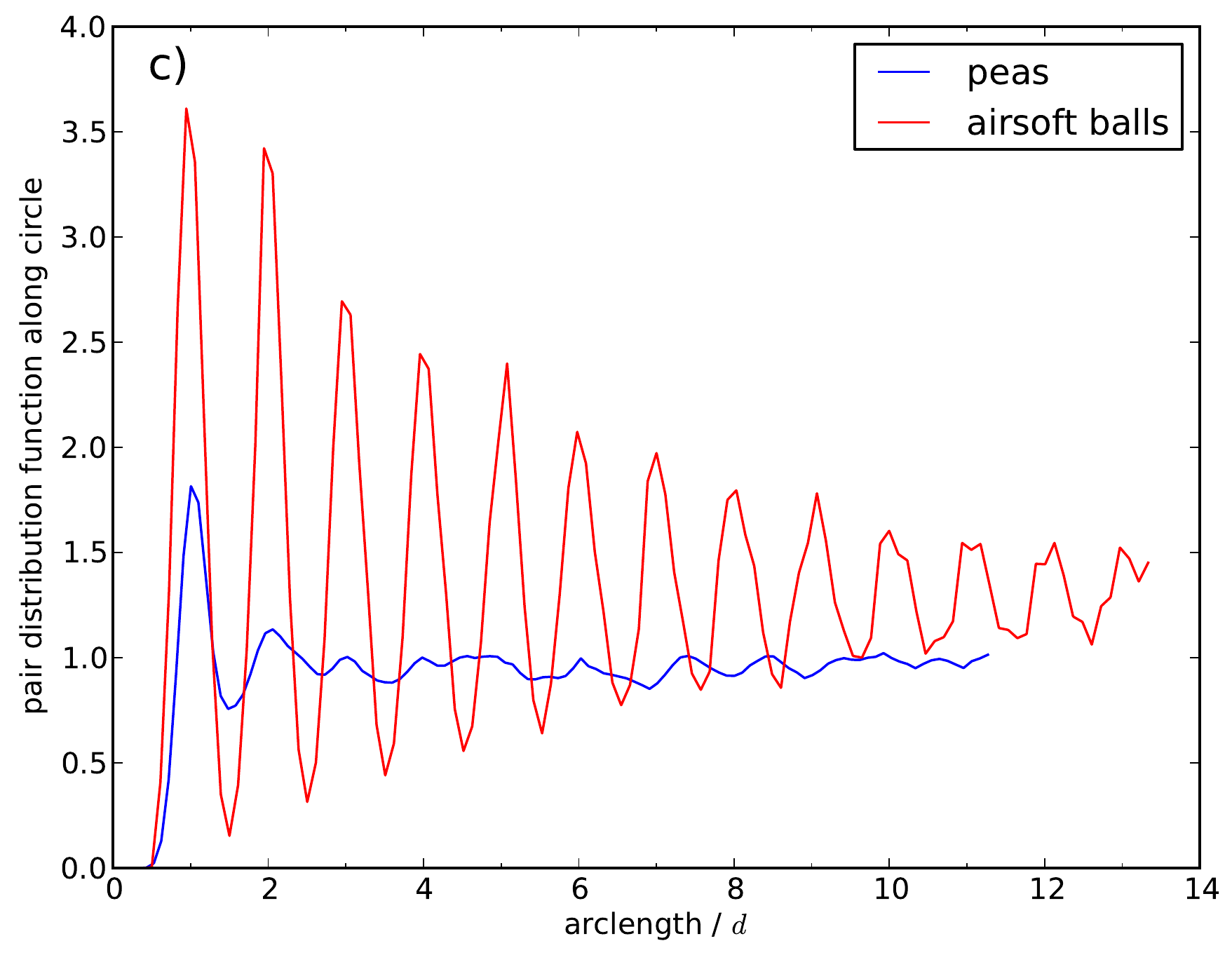}}
	\caption{Pair distribution functions for (a) peas and (b) airsoft balls
in the asymptotic state in the center of the zone, obtained by particle
detection.  The functions show the probability of finding a pair of particle
centers $\Delta x', \Delta y', \Delta z'$ apart, normalised by the probability
expected for a completely random distribution of particles at the same
packing density \cite{allentildesley}.  $x'$ is in
the radial direction for the first particle, $y'$ is the perpendicular
direction in the horizontal plane, and the height difference is restricted to
$|\Delta z'| < d/2$.
(c) The pair distribution function along the streamlines (the black lines
indicated on panels a-b).  For peas little structure is observed beyond the
first neighbourhood shell of the hard-core particles, as expected from the lack
of spatial patterns (upper image, Fig.\ \ref{fig:ellak2}).  For the regular
airsoft balls, however, a strong periodicity is observed along the curved
chains following the streamlines, consistent with the spatial patterns seen on
the upper image of Fig.\ \ref{fig:ellak1}.  
	\label{fig:pair}}
\end{figure*}

The pair distribution functions (obtained by particle detection) have been calculated for particles in the center 
of the zone for the asymptotic state.
As it is seen in Fig. \ref{fig:pair} for peas the distribution of the first neigbours at a distance $d$ is about isotropic,
no strong long distance correlations are visible,
and the pair distribution function quickly converges to 1 as expected. On the other hand for airsoft balls the particles clearly line up
in the flow direction, where one can identify even the average position of the 13th neigbour. The curved trajectories 
representing 
the chains sliding next to each other (see Figs. \ref{fig:ellak1} and \ref{fig:hexagons-lattice}a) are at a well defined 
distance from each other, which is sligtly smaller than $d$, as it was also indicated in Fig. \ref{fig:hexagons-lattice}b.
The pair distribution function for airsoft balls along the streamline does not converge to 1, indicating the long
range correlations in the chain-like structures.
(The pair distribution function converges to 1 at large distances
when averaged over regions containing both chains and inter-chain areas.)

%%% FIG 8 %%%%%%%%%%%%%%%%%%%%%%%%%%%%%%%%%%%%%%%%%%%

\subsection{Dilation}
\label{Dilation}

%%% FIG 9 %%%%%%%%%%%%%%%%%%%%%%%%%%%%%%%%%%%%%%%%%%%
\begin{figure}[htbp]
\centerline{\includegraphics[width=\columnwidth,clip]{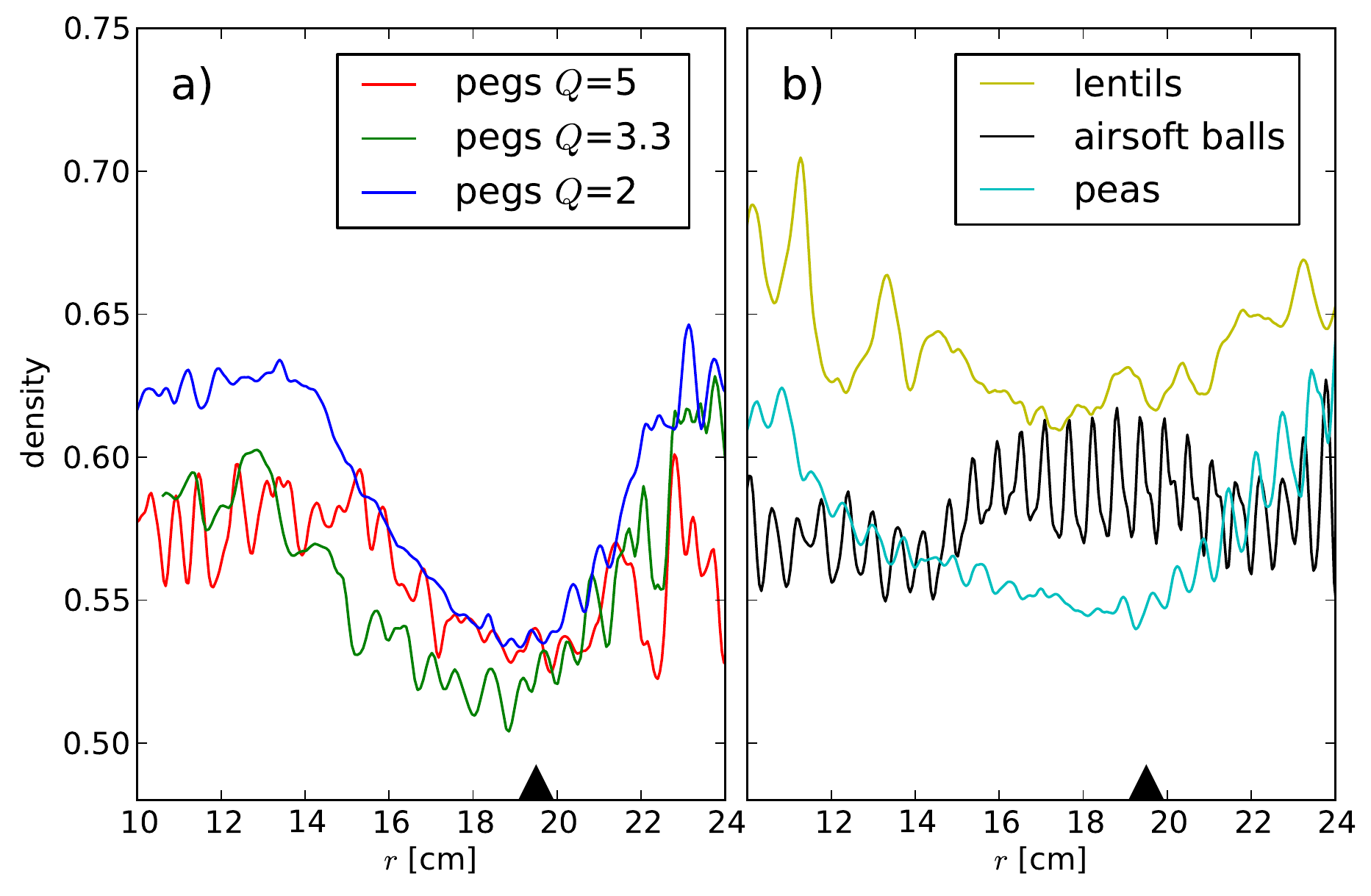}}
	\caption{Averaged packing density of grains of different shapes near the shear zone. The densities
shown in Fig.~\ref{fig:dichten_alle} were averaged in a 1 cm thick horizontal layer below the surface.
The black triangle marks the position of the edge of the rotating plate.
	\label{fig:mittelung1}}
\end{figure}
The macroscopic consequences of positional and orientational order on packing densities can be expressed in a more
quantitative way:
The mean packing densities in different zones of the sheared material are
obtained by averaging over the data shown in Fig.~\ref{fig:dichten_alle}a-f.
We average over a representative 1 cm thick layer not far (typically between 1 and 2 cm) below the initial top surface
of the sample.

It is evident that in all samples except the perfect spheres, the packing density decreases inside the
shear zone.
The amplitude of the relative decrease is between 9-16 $\%$. Interestingly, the strongest decrease is observed for
one of the pegs (with $Q=2$), even though one would intuitively assume that for non-spherical grains the shear 
induced ordering partially compensates shear dilation. The larger the aspect ratio $Q$ of elongated grains, 
the weaker are the dilatancy effects (Fig.~\ref{fig:mittelung1}a), which is in accordance with the fact, that 
shear induced ordering is more pronounced for longer particles \cite{Borzsonyi2012}.

The other striking effect is observed for the case of uniform airsoft balls, where shear induced layering not only compensates,
but also overcomes the effect of shear dilation, leading to slightly (about 5 $\%$) larger average density in the shear zone
than elsewhere.
For the case of nearly spherical grains (peas) we find a density decrease of about 11$\%$, which is consistent
with dilation of poppy seeds measured by Magnetic Resonance Imaging \cite{Sakaie2008}.
When comparing the average packing densities for different materials outside the zone we find, that according to
the expectations  \cite{Donev2004,Donev2005} random packing density of pegs is decreasing when the aspect ratio
of the grains is increased from 2 to 5. Also, slightly nonuniform beads (peas) form a denser randomly packed system, than the uniform airsoft balls.

\subsection{Depression of the shear zone surface}

\begin{figure}[htbp]
	\includegraphics[width=\columnwidth,clip]{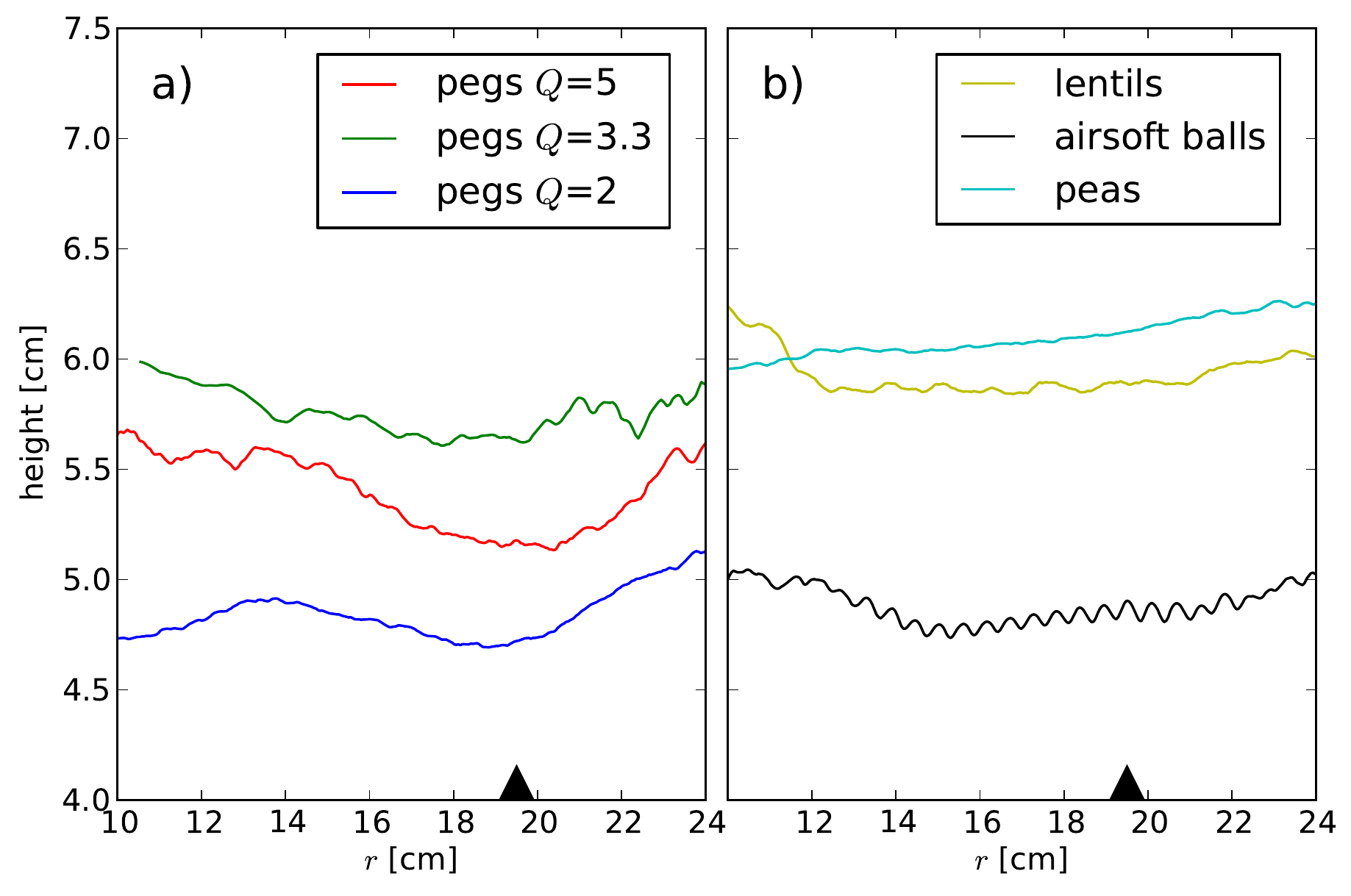}
	\caption{Height of the granular bed across the shear zone for (a) pegs with $Q = 2.0$, 3.3 and 5.0, and (b) lentils, peas and airsoft balls. The grey triangle indicates the edge of the bottom plate.
 }
	\label{fig:surface}
\end{figure}

After filling, the granular bed in the container has a flat surface. As is seen from Figure~\ref{fig:dichten_alle}, the surface does not remain flat for all materials. Averaged surface profiles for the different materials are plotted in Fig.~\ref{fig:surface}. The height profile of the surface above the shear zone depends significantly upon the shape of the material. While the nearly spherical peas and the oblate ellipsoid lentils preserve a nearly flat profile, the prolate cylinders develop a pronounced depression of the height profile above the shear zone. For the regular spheres, the height profile is discrete owing to the crystalline lattice, but the modulation of the height profile does not exceed one layer.

The depression of the granular surface above the shear zone for the prolate objects is accompanied by heaping in the
adjacent regions. On one hand, we assume that one reason for the development of the heaps is the non-trivial time
evolution, where the system first expands and a certain amount of local strain is needed for starting the
compaction (for details see next section). On the other hand, it is also possible that a slow material flow develops
normal to the flow direction. As has been shown earlier (Ref.~\cite{Borzsonyi2012a}, Figs.~7-9), the elongated grains
perform a continuous reorientation in the shear zone, with angular velocities that depend upon their orientation
respective to the flow. This permanent reorientation could create a constant pressure that drives material out of
the shear zone, which could lead to heaping in the adjacent regions. However, we cannot provide direct experimental 
evidence for this convection with the experiments presented here.

%%% FIG 10 %%%%%%%%%%%%%%%%%%%%%%%%%%%%%%%%%%%%%%%%%%%

\subsection{Temporal Evolution}

\label{sec:time}

\begin{figure}[htbp]%
        \includegraphics[width=\columnwidth,clip]{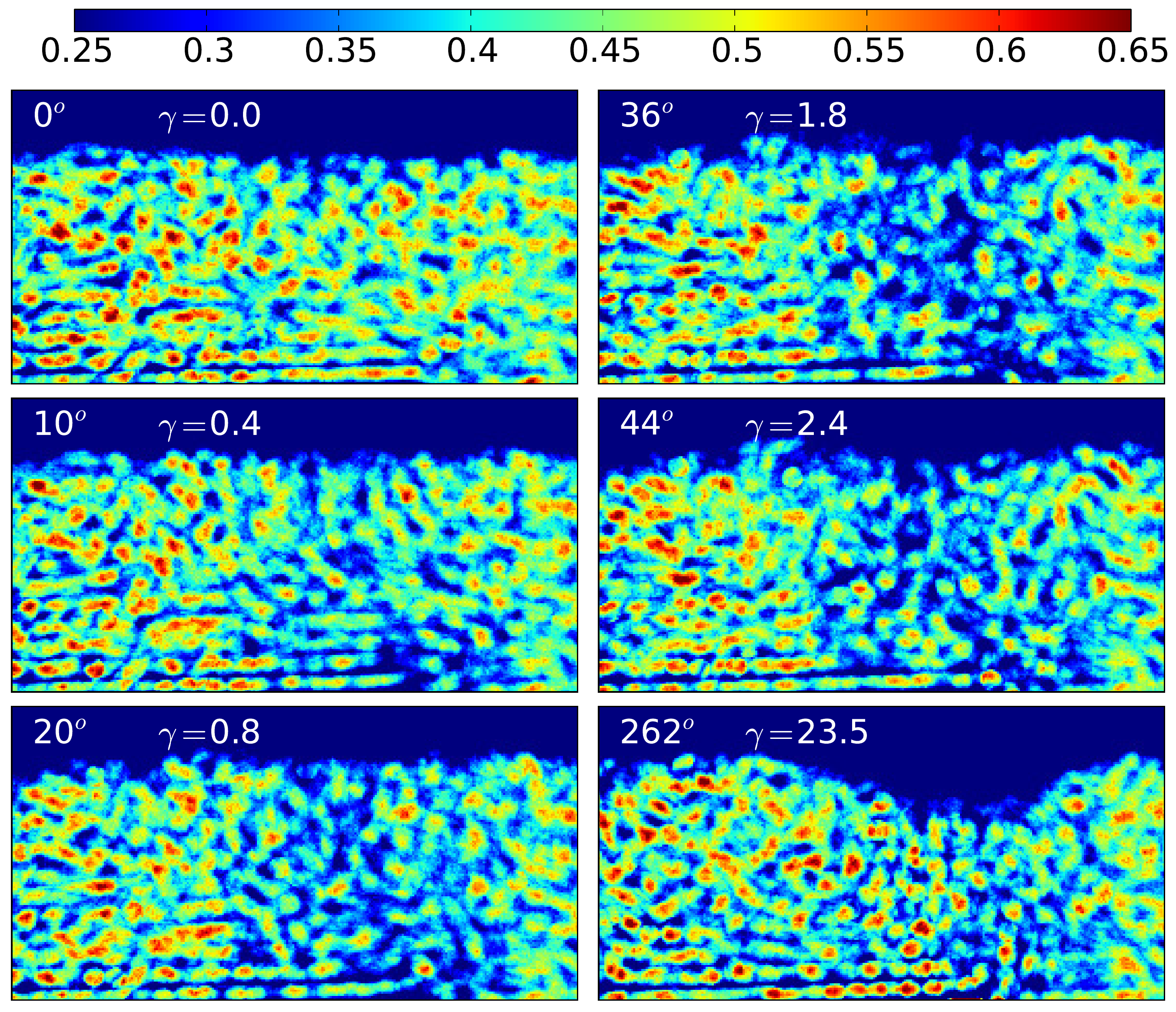}

        \caption{Evolution of the averaged packing densities of the $Q=5$ pegs, obtained from the angular averaging 
(Fig.~\ref{fig:dichten_alle}g) of single tomograms taken in  2$^\circ$  rotation steps. 
One acknowledges three phases: first the dilation in the initial stage, which leads to heaping above the shear zone
as an effect of the reduced packing density of the grains below. Second, the heap levels and transforms into a depression
above the core of the shear zone since local orientational ordering  first develops here.
Afterwards, the depression spreads outwards as the grains become ordered throughout the shear zone.
The images were obtained with the same material as in Fig.~\ref{fig:dichten_alle}a, but in a different experimental run.
For each frame the rotation angle of the inner disk is indicated together with the maximal local strain $\gamma$.
The fact that $\gamma$ is not strictly linear with the rotation angle is attributed to the time evolution of the 
zone width during the initial transient, which will be discussed in details elsewhere \cite{Szabo2014}.
}
        \label{fig:timeQ2}
\end{figure}

\begin{figure}[htbp]%
        \includegraphics[width=\columnwidth]{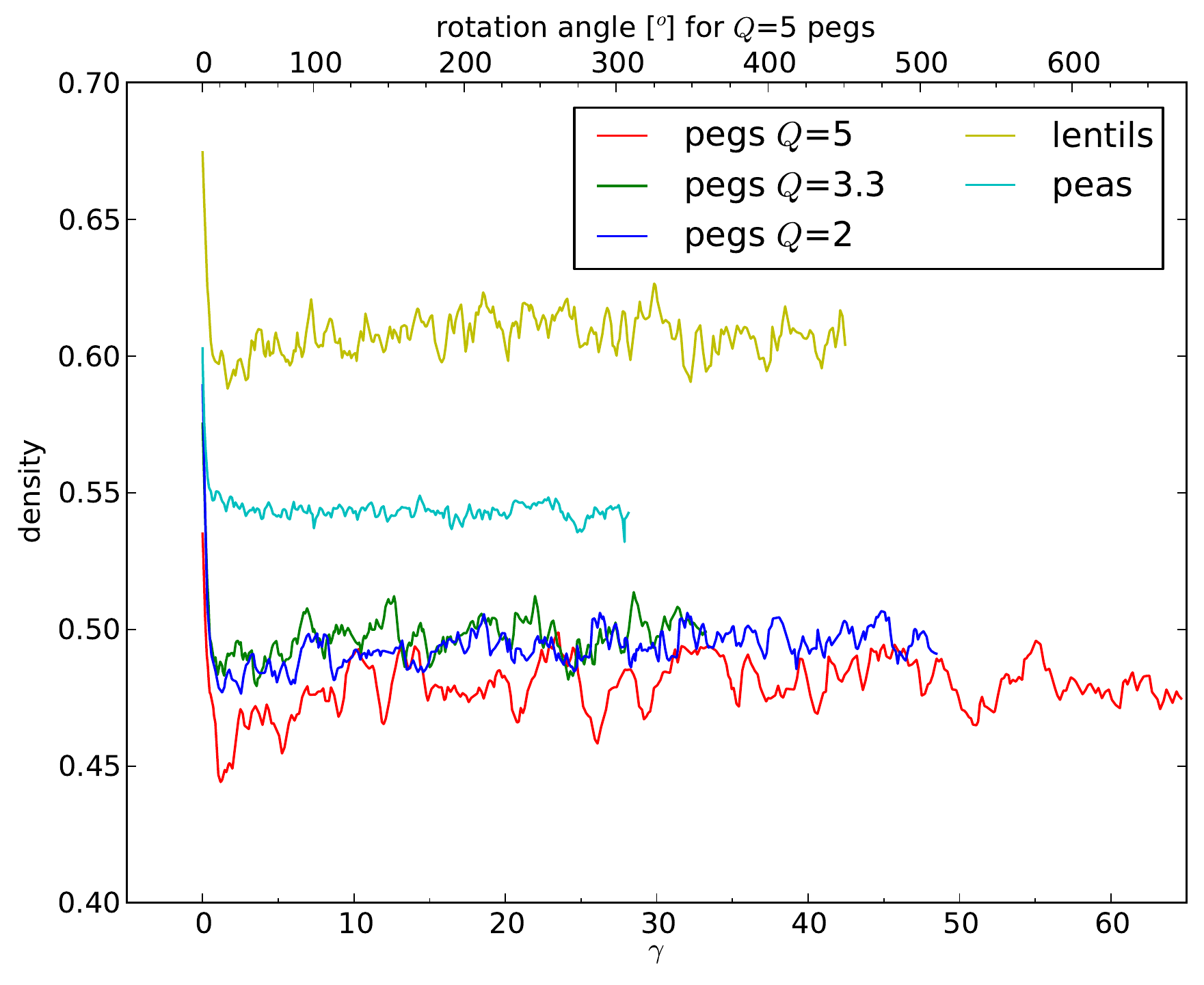}
        \caption{Development of the packing density as a function of the maximal local strain $\gamma$ in a
representative (typically 4 cm thick) region of the shear zone, where a 1 cm thick layer below the initial
surface and a 1 cm thick layer above the rotating plate were excluded.
A rapid decrease observed for all materials is attributed to dilatancy. For shape anisotropic grains (pegs and lentils)
this is followed by a partial, slower recovery, related to the orientational ordering of the material.
The given rotation angle of the bottom plate (top axis) corresponds to the case of $Q=5$ pegs.}
        \label{fig:develop}
\end{figure}

Two time scales (i.e. shear strain scales) are involved in the evolution of positional or orientational order and dilatancy.
The transient behavior of the freshly prepared granular bed is recorded by tomograms taken in 2$^\circ$ rotation steps.
We are particularly interested in the evolution of the packing density and the surface structure
in elongated grains. Representative images are shown in Fig.~\ref{fig:timeQ2} for the longest ($Q=5$) pegs.
%where the most pronounced effects have been found in the asymptotic state.
For the other cylinders, the results are
qualitatively similar. The packing density quickly drops in the shear zone after the start of shearing.
%
%%% FIG 11 %%%%%%%%%%%%%%%%%%%%%%%%%%%%%%%%%%%%%%%%%%%
To quantify this scenario, the evolution of the local density in the core of the zone has been measured for 5 samples and
it is shown in Fig.~\ref{fig:develop} as a function of the maximal local strain $\gamma$, where the value of $\gamma$ has 
been determined from the particle displacements (as described in refs. \cite{Borzsonyi2012,Borzsonyi2012a}).

The rapid density decrease is clearly seen for all 5 samples in the beginning of the process. 
%
%%% FIG 12 %%%%%%%%%%%%%%%%%%%%%%%%%%%%%%%%%%%%%%%%%%%
This dilation corresponds to about 17 \% for the case of pegs with $Q=5$.
The rotation angle ($20^\circ$) at this point approximately corresponds to a total maximal local strain $\gamma$ of the 
order of one (i.e. particles have been displaced by about one particle diameter with respect to their neighbors). 
The reduced packing density results in an
expansion of the material, thus an elevation of the surface forms (Fig.~\ref{fig:timeQ2}, left). This affects a wider
region than the density decrease, i.e. the material is pushed not exclusively upwards but sideways as well.
For the case of pegs and lentils this elevation collapses as shear induced orientational ordering and the corresponding local
increase of the density develops. This happens first above the core of the zone (Fig.~\ref{fig:timeQ2}, right middle).
For pegs with $Q=5$, the local density increases back by about 7 \% (see Fig.~\ref{fig:develop}), the
local strain needed for this process is about $\gamma=5$ to $10$, corresponding to rotation angles of $70$ to 
$140^\circ$.
Thereafter, the surface above the shear zone gradually sinks further, as shear induced ordering develops in the whole 
zone. After an approximately three-quarters turn of the center, a depression of the upper surface has formed 
(Fig.~\ref{fig:timeQ2}, bottom right). 
Note that for the peas there is no density increase at any time. 

%
%%% FIG 13 %%%%%%%%%%%%%%%%%%%%%%%%%%%%%%%%%%%%%%%%%%%
\begin{figure}[htbp]%
        \includegraphics[width=\columnwidth]{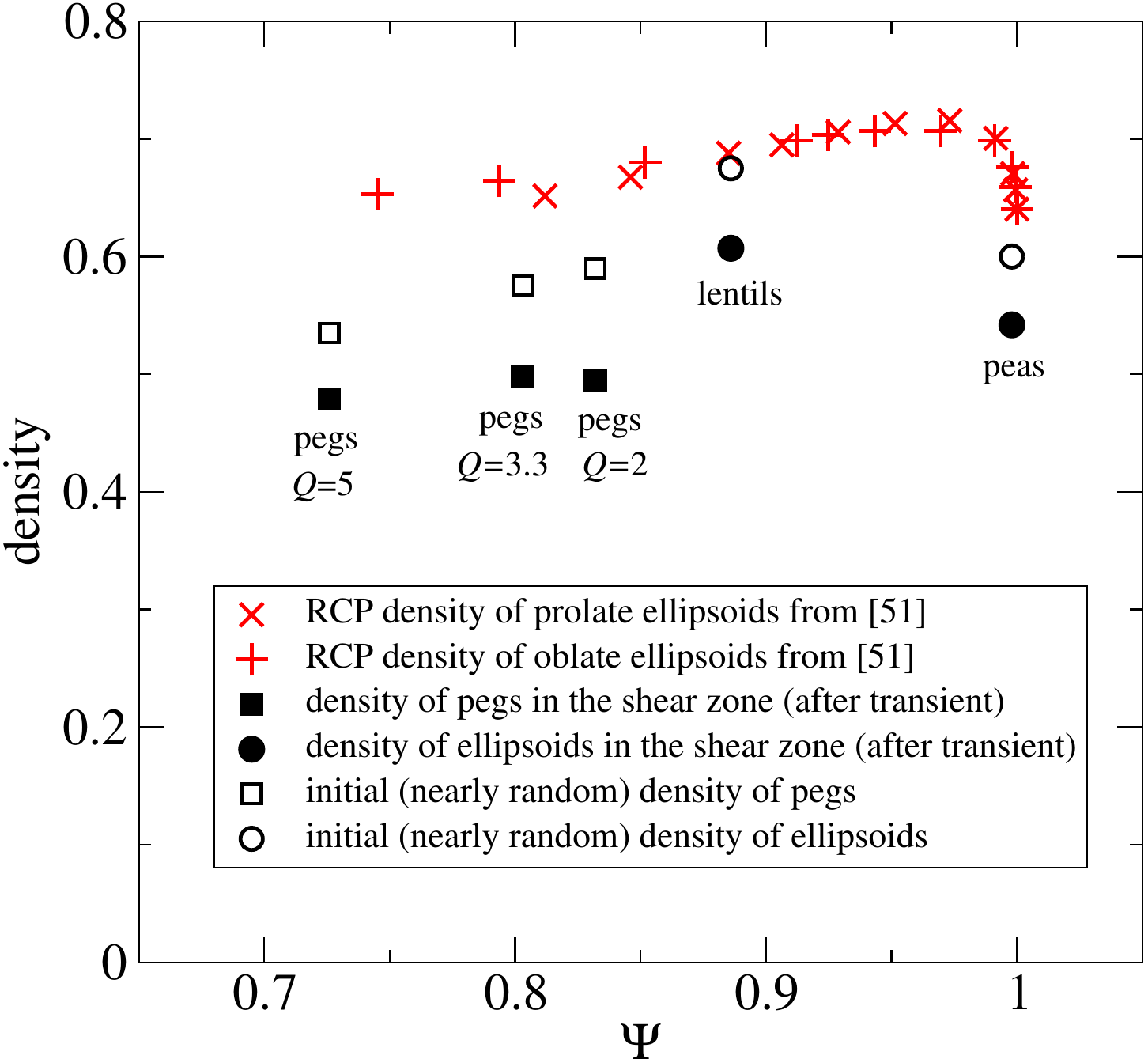}
        \caption{The packing density as a function of the sphericity $\Psi$ of the particles in the initially (nearly) 
random configuration and in the shear zone after the initial transient. 
Random closed packed (RCP) densities obtained for prolate and oblate ellipsoids in Ref. \cite{Donev2004} 
are also included for comparison.}
        \label{fig:sphericity-density}
\end{figure}

Previous studies suggested that various parameters of a sheared granular assembly depend systematically on the sphericity 
of the particles \cite{Siang2013,Xu2013,Xu2014} (as a general measure of the deviation from the spherical shape). 
Our data in Fig.  \ref{fig:sphericity-density} show the initial (nearly random) packing density as well as the density 
in the shear zone after the transient for pegs with 3 different aspect ratio, lentils and peas as a function of 
sphericity $\Psi$.
In this graph, we have also included the random closed packed (RCP) density of ellipsoids taken from Ref. \cite{Donev2004}
which include both branches, prolate and oblate ellipsoids, both showing a clear non-monotonous tendency with sphericity.
Both of our datasets corresponding to the initial (nearly random) packing and data from the shear zone are consistent 
with this tendency. Our initial (nearly random) packing density data are somewhat below the RCP curves for ellipsoids,
as we are somewhere between the RCP and random loose packed (RLP) density. The rod like shape of pegs leads to an even 
smaller initial density than what one would expect for ellipsoids. Altogether we clearly see a downward shift of the 
data points as a consequence of the shear induced dilation.

%---------------------------------------------------------------------------------------------------------
\section {Conclusions and Summary}

The influence of shape and polydispersity of granular materials on their dilatancy under shear has been studied both on the
individual grain level and on the macroscopic level. We find a competition of dilatancy effects with orientational and positional ordering
that both tend to increase the packing density of grains in the shear zone. The following observations were made:

(a) Positional ordering of perfectly spherical, monodisperse grains in hexagonally stacked chains aligned to the shear
direction is very efficient for the compensation of dilatancy. In fact, the packing density increases in the shear zone 
with respect to randomly packed grains outside by approximately 5~\%.

(b) The hexagonally packed chains of spheres follow the curved flow direction.

(c) The crystallization extends far into the regions outside the kernel of the shear zone. In these regions, the spheres 
develop a positional order without shear, i.e. they crystallize even though they are displaced respective to their 
neighbors by less than the particle size. The reason for that are small creeping motions of the granulate outside the 
zone where shear is focused.

(d) Small deviations from spherical shape and/or a slight polydispersity of the grains completely inhibits the 
crystallization. It was not possible within this study to distinguish whether the deviations from exact sphere 
shape or the polydispersity alone or a combination of both cause this inhibition.
Sheared peas show a conventional dilatancy, and almost no regular positional order except some vertical layering. This layering is particularly pronounced at the container bottom (see Fig~\ref{fig:dichten_alle}f), and it decays with increasing distance to the bottom plate. It can therefore be attributed to boundary effects.

(e) Anisometric grains (cylinders and oblate ellipsoids) as well as nearly spherical particles (peas)
show dilatancy of comparable magnitude ($9-16\%$) in the shear zone. Interestingly, the strongest dilation is observed
for the case of cylinders even if the shear induced expansion of the material is partially compensated by orientational
ordering of the grains. For the elongated grains, increasing aspect ratio $Q$ leads to weaker dilatancy,
which is in accordance with the fact, that shear induced ordering (the compensating effect)
is more pronounced for longer particles \cite{Borzsonyi2012}.

(f) The time evolution of the packing structure of sheared cylinders is divided into three phases:
First, the dilation of the shear zone sets in, which leads to an extension of the granular bulk and formation
of an elevated surface which actually affects a wider region than the dilation, as the material is pushed not
exclusively upwards but sideways as well. Second, a depression of the surface develops above the core of the 
zone where an induced order develops first. 
Finally, the depression extends further away from the core of the zone, as orientational ordering and the corresponding
compaction gradually develops throughout the whole shear zone.

\section{Acknowledgements}
Financial support by the DAAD/M\"OB researcher exchange program (grant no. 29480), the Hungarian Scientific Research Fund 
(grant no. OTKA NN 107737), and the J\'anos Bolyai Research Scholarship of the Hungarian Academy of Sciences is acknowledged.
We kindly acknowledge Piasten GmbH $\&$ Co. KG for supplying us with chocolate lentils for the X-ray CT study.

\footnotesize{
\bibliography{packung}
\bibliographystyle{rsc}
}

\end{document}